\documentstyle[preprint,aps,prd,eqsecnum,twoside]{revtex}
{\hspace*{\fill}{\protect\small 
{\bf Bijan~Saha}}
\hspace*{\fill}
}
{\hspace*{\fill}
{\protect\small {\bf Nonlinear Spinor Field in Anisptropic 
Universes}}
\hspace*{\fill}
}
\newcommand {\cG}{\cal G}
\newcommand {\cD}{\cal D}
\newcommand {\bg}{\bar \gamma}
\newcommand {\bp}{\bar \psi}
\newcommand {\bv}{\bar v}
\newcommand {\p}{\psi}
\begin{document}
\title{Nonlinear Spinor Field in Anisotropic Universes\footnote{
Talk, given at the International Conference ``Scientific Reading
devoted to 90 years anniversary of Professor Yakov Petrovich 
Terletskii'', July 1-3, 2002, Russian Peoples' Friendship
University, Moscow, Russia.}}
\author{Bijan Saha \\
{\small \it Laboratory of Information Technologies}\\
{\small \it Joint Institute for Nuclear Research}\\
{\small \it 141980 Dubna, Moscow Reg., Russia.\\
e-mail: saha@thsun1.jinr.ru}}
\maketitle
\begin{abstract}
Evolution of an anisotropic universe described by a Bianchi
type I (BI) model in presence of nonlinear spinor field has
been studied by us recently in a series of papers. On offer
the Bianchi models, those are both inhomogeneous and
anisotropic. Within the scope of Bianchi type VI (BVI) model
the self-consistent system of nonlinear spinor and gravitational 
fields are considered. The role of inhomogeneity in the evolution
of spinor and gravitational field is studied. 

\end{abstract}
\section{Introduction}
\setcounter{equation}{0}
Several authors studied the nonlinear spinor fields (NLSF) 
since Ivanenko \cite{ivanenko1,ivanenko2,rodichev} showed that 
a relativistic theory imposes in some cases a fourth order
self-coupling. Nonlinear spinor field in an anisotropic universe,
namely in a Bianchi-type I universe (BI) is studied by us in
a series of papers~
\cite{sahapfu1,sahactp1,sahajmp,sahactp2,sahaizv,sahagrg,pfu-l,sahal}.
In these papers we considered the nonlinear spinor field, as well as
a system of interacting spinor and scalar fields. Beside the 
spinor fields, we also study the role of a $\Lambda$ term in the 
evolution of the universe. For the details of a NLSF in a BI universe
one can consult~\cite{sahaprd}. In the light of results obtained
in the previously mentioned papers the study of NLSF in other anisotropic
universes presents great interest. In this report we consider the  
self-consistent system of nonlinear spinor and BVI gravitational 
fields. The results have been compared with those obtained for the
BI universe. 

\noindent
\vskip 5mm
\section{Fundamental equations and general solutions}
We choose the Lagrangian for the self-consistent system of  spinor  and 
gravitational fields in the form 
\begin{equation} 
L=\frac{R}{2\kappa}+\frac{i}{2} 
\biggl[ \bp \gamma^{\mu} \nabla_{\mu} \p- \nabla_{\mu} \bp 
\gamma^{\mu} \p \biggr] - M\bp \p + L_N  \label{lag}
\end{equation} 
with $R$ being the scalar curvature and $\kappa$ being  the  
Einstein's gravitational constant. The nonlinear term
$L_N$ describes the self-interaction of spinor field and can be presented
as some arbitrary functions of invariants generated from the real bilinear 
forms of spinor field having the form: 
\begin{mathletters}
\label{bf}
\begin{eqnarray}
 S&=& \bar \psi \psi\qquad ({\rm scalar}),   \\                   
  P&=& i \bar \psi \gamma^5 \psi\qquad ({\rm pseudoscalar}), \\
 v^\mu &=& (\bar \psi \gamma^\mu \psi) \qquad ({\rm vector}),\\
 A^\mu &=&(\bar \psi \gamma^5 \gamma^\mu \psi)\qquad ({\rm pseudovector}), \\
Q^{\mu\nu} &=&(\bar \psi \sigma^{\mu\nu} \psi)\qquad
({\rm antisymmetric\,\,\, tensor}),  
\end{eqnarray}
\end{mathletters}
where $\sigma^{\mu\nu}\,=\,(i/2)[\gamma^\mu\gamma^\nu\,-\,
\gamma^\nu\gamma^\mu]$. 
Invariants, corresponding to the bilinear forms, are
\begin{mathletters}
\label{invariants}
\begin{eqnarray}
I &=& S^2, \\
J &=& P^2, \\ 
I_v &=& v_\mu\,v^\mu\,=\,(\bar \psi \gamma^\mu \psi)\,g_{\mu\nu}
(\bar \psi \gamma^\nu \psi),\\ 
I_A &=& A_\mu\,A^\mu\,=\,(\bar \psi \gamma^5 \gamma^\mu \psi)\,g_{\mu\nu}
(\bar \psi \gamma^5 \gamma^\nu \psi), \\
I_Q &=& Q_{\mu\nu}\,Q^{\mu\nu}\,=\,(\bar \psi \sigma^{\mu\nu} \psi)\,
g_{\mu\alpha}g_{\nu\beta}(\bar \psi \sigma^{\alpha\beta} \psi). 
\end{eqnarray}
\end{mathletters}

According to the Pauli-Fierz theorem \cite{Ber} among the five invariants
only $I$ and $J$ are independent as all other can be expressed by them:
$I_V = - I_A = I + J$ and $I_Q = I - J.$ Therefore, we choose the nonlinear
term $L_N = F(I, J)$, thus claiming that it describes the nonlinearity
in the most general of its form. 

We choose the anisotropic inhomogeneous universe given by a
Bianchi-type VI (BVI) metric:
\begin{equation} 
ds^2 = dt^2 - a^{2} e^{-2mz}\,dx^{2} - b^{2} e^{2nz}\,dy^{2} - 
c^{2}\,dz^2, 
\label{met}
\end{equation}
with $a,\,b,\,c$ being the functions of time only. Here 
$m,\,n$ are some arbitrary constants and the velocity of light is 
taken to be unity. Note that the suitable choice of $m,\,n$ as well 
as the metric functions $a,\,b,\,c$ in the BVI given by (\ref{met}) 
evokes the following Bianchi-type universes. Thus
\begin{itemize}
\item
for $m=n$ the BVI metric transforms to a Bianchi-type V (BV) one, i.e.,\\
$m = n$, BVI $\Longrightarrow$ BV $\in$ open FRW;
\item
for $n=0$ the BVI metric transforms to a Bianchi-type III (BIII) one, i.e.,\\
$n = 0$, BVI $\Longrightarrow$ BIII;
\item
for $m=n =0$ the BVI metric transforms to a Bianchi-type I (BI) one, i.e.,\\
$m = n = 0$, BVI $\Longrightarrow$ BI;
\item
for $m=n=0$ and equal scale factor in all three direction the BVI metric 
transforms to a Friedmann-Robertson-Walker (FRW) universe, i.e.,\\
$m = n = 0$ and $a=b=c$, BVI $\Longrightarrow$ FRW.
\end{itemize}

Variation of the Lagrangian (\ref{lag}) with respect to 
field functions $\psi (\bp)$  gives the nonlinear spinor field 
equations:
\begin{mathletters}
\label{spin}
\begin{eqnarray}
i\gamma^\mu \nabla_\mu \p -M\p + {\cD} \p + i {\cG} \gamma^5 \p &=& 0 \\
i \nabla_\mu \bp \gamma^\mu + M \bar\psi - {\cD} \bp - 
i {\cG} \bp \gamma^5 &=& 0 
\end{eqnarray}
\end{mathletters}
where ${\cD} = 2 S F_I$ and ${\cG} = 2 P F_J$.  

Varying (\ref{lag}) with respect to metric function ($g_{\mu\nu}$)
we find the Einstein equations for $a,\,b,\,c$ which for the 
 metric~(\ref{met}) read  
\begin{mathletters}
\label{ein}
\begin{eqnarray}
\frac{\ddot b}{b} +\frac{\ddot c}{c} +\frac{\dot b}{b}\frac{\dot 
c}{c} - \frac{n^2}{c^2} &=& -\kappa T_{1}^{1}, \label{11}\\
\frac{\ddot c}{c} +\frac{\ddot a}{a} +\frac{\dot c}{c}\frac{\dot 
a}{a} - \frac{m^2}{c^2} &=& -\kappa T_{2}^{2}, \label{22} \\
\frac{\ddot a}{a} +\frac{\ddot b}{b} +\frac{\dot a}{a}\frac{\dot 
b}{b} + \frac{m n}{c^2} &=& -\kappa T_{3}^{3}, \label{33}\\
\frac{\dot a}{a}\frac{\dot b}{b} +\frac{\dot b}{b}\frac{\dot c}{c} +
\frac{\dot c}{c}\frac{\dot a}{a} - \frac{m^2 - m n + n^2}{c^2} &=& 
-\kappa T_{0}^{0}, \label{00}\\
m \frac{\dot a}{a} - n \frac{\dot b}{b}
- (m - n) \frac{\dot c}{c} &=& -\kappa T_{3}^{0}. \label{03}
\end{eqnarray}
\end{mathletters}
Here overdots denote differentiation with respect to time ($t$) and 
$T_{\mu}^{\nu}$ is the energy-momentum tensor of the material 
field and has the form
\begin{equation}
T_{\mu}^{\rho}=\frac{i}{4} g^{\rho\nu} \biggl(\bp \gamma_\mu 
\nabla_\nu \psi + \bp \gamma_\nu \nabla_\mu \psi - \nabla_\mu \bp 
\gamma_\nu \psi - \nabla_\nu \bp \gamma_\mu \psi \biggr) \,-
\delta_{\mu}^{\rho}L_{sp}.
\label{emt}
\end{equation}
Here $L_{sp}$ is the spinor field Lagrangian, which 
on account of spinor field equations~(\ref{spin}) takes the form:
\begin{equation}
L_{sp} = - {\cD} S - {\cG P} + F.
\label{sflag}
\end{equation}
In the expressions above $\nabla_\mu$ denotes the covariant 
derivative of spinor, having the form \cite{Zelnor}:  
\begin{equation} \nabla_\mu 
\p=\partial_\mu \p - \Gamma_\mu \p \end{equation} 
where $\Gamma_\mu(x)$ are spinor affine connection matrices 
defined by the equality 
$$\Gamma_\mu (x)= 
(1/4) g_{\rho\sigma}(x)\bigl(\partial_\mu e_{\delta}^{b}e_{b}^{\rho} 
- \Gamma_{\mu\delta}^{\rho}\bigr)\gamma^\sigma\gamma^\delta. $$ 
For the metric element (\ref{met}) it gives
\begin{eqnarray} 
\Gamma_0 &=& 0, \nonumber\\ 
\Gamma_1 &=& \frac{1}{2}\bigl[{\dot a} \bg^{1} \bg^0
- m \frac{a}{c} \bg^{1} \bg^3 \bigr] e^{-mz} \nonumber \\  
\Gamma_2 &=& \frac{1}{2}\bigl[ {\dot b} \bg^{2} \bg^0
+ n \frac{b}{c} \bg^{2} \bg^3 \bigr] e^{nz}, \nonumber\\
\Gamma_3 &=& \frac{1}{2} {\dot c} \bg^{3} \bg^0  \nonumber
\end{eqnarray}
It is easy to show that
$$\gamma^\mu \Gamma_\mu = -\frac{1}{2}\frac{\dot \tau}{\tau}\bg^0
+ \frac{m - n}{2 c}\bg^3,$$
where we define 
\begin{equation}
\tau = a b c.
\label{taudef}
\end{equation}
The Dirac matrices $\gamma^\mu(x)$ of curved space-time are connected
with those of Minkowski one as follows:
$$ \gamma^0=\bg^0,\quad \gamma^1 =\bg^1 e^{mz}/a,
\quad \gamma^2=\bg^2 /b e^{nz},\quad \gamma^3 =\bg^3 /c$$
with 
\begin{eqnarray}
\bg^0\,=\,\left(\begin{array}{cc}I&0\\0&-I\end{array}\right), \quad
\bg^i\,=\,\left(\begin{array}{cc}0&\sigma^i\\
-\sigma^i&0\end{array}\right), \quad
\gamma^5 = \bg^5&=&\left(\begin{array}{cc}0&-I\\
-I&0\end{array}\right),\nonumber
\end{eqnarray}
where $\sigma_i$ are the Pauli matrices:
\begin{eqnarray}
\sigma^1\,=\,\left(\begin{array}{cc}0&1\\1&0\end{array}\right), \quad
\sigma^2\,=\,\left(\begin{array}{cc}0&-i\\i&0\end{array}\right), \quad
\sigma^3\,=\,\left(\begin{array}{cc}1&0\\0&-1\end{array}\right).\nonumber
\end{eqnarray}
Note that the $\bg$ and the $\sigma$ matrices obey the following properties:
\begin{eqnarray}
\bg^i \bg^j + \bg^j \bg^i = 2 \eta^{ij},\quad i,j = 0,1,2,3 \nonumber\\
\bg^i \bg^5 + \bg^5 \bg^i = 0, \quad (\bg^5)^2 = I, \quad i=0,1,2,3 \nonumber\\
\sigma^j \sigma^k = \delta_{jk} + i \varepsilon_{jkl} \sigma^l, \quad
j,k,l = 1,2,3 \nonumber
\end{eqnarray}
where $\eta_{ij} = \{1,-1,-1,-1\}$ is the diagonal matrix, $\delta_{jk}$
is the Kronekar symbol and $\varepsilon_{jkl}$ is the totally antisymmetric 
matrix with $\varepsilon_{123} = +1$.
Let us consider the spinors to be functions of $t$ and $z$ only, such that
\begin{equation}
\p (t, z) = v (t) e^{i k z}, \quad \bp (t,z) = \bv (t) e^{- i k z}
\label{z}
\end{equation}

Inserting (\ref{z}) into (\ref{spin}) for the nonlinear spinor field 
we find
\begin{mathletters}
\label{spinv}
\begin{eqnarray}
\bg^0\Bigl(\dot v + \frac{\dot \tau}{2 \tau} v\Bigr) 
-\Bigl(\frac{m-n}{2 c} - i\frac{k}{c}\Bigr)\bg^3 v + i\Phi v 
+ {\cG} \bg^5 v &=& 0 \\
\Bigl(\dot{\bv} + \frac{\dot \tau}{2 \tau}\bv\Bigr)\bg^0 
-\Bigl(\frac{m-n}{2 c} + i\frac{k}{c}\Bigr) \bv \bg^3 - i\Phi \bv  
- {\cG} \bv \bg^5  &=& 0. 
\end{eqnarray}
\end{mathletters}
Here we define $\Phi = M - {\cD}$.
Introduce a new function $u_j (t) = \sqrt{\tau} v_j(t)$, 
for the components of the NLSF from (\ref{spinv}) one obtains
\begin{mathletters}
\label{u}
\begin{eqnarray} 
\dot{u}_{1} + i \Phi u_{1} - 
\Bigl[\frac{m-n}{2 c} - i\frac{k}{c} +{\cG}\Bigr] u_{3} &=& 0, \\
\dot{u}_{2} + i \Phi u_{2} + 
\Bigl[\frac{m-n}{2 c} - i\frac{k}{c} -{\cG}\Bigr] u_{4} &=& 0, \\
\dot{u}_{3} - i \Phi u_{3} - 
\Bigl[\frac{m-n}{2 c} - i\frac{k}{c} -{\cG}\Bigr] u_{1} &=& 0, \\
\dot{u}_{4} - i \Phi u_{4} + 
\Bigl[\frac{m-n}{2 c} - i\frac{k}{c} +{\cG}\Bigr] u_{2} &=& 0. 
\end{eqnarray} 
\end{mathletters}
Using the spinor field equations (\ref{spin}) and (\ref{spinv})
it can be shown that the bilinear spinor forms, defined
by (\ref{bf}), i.e.,  
\begin{eqnarray}
S &=& \bp \p = \bv v, \quad            
P = i \bp \bg^5 \p = i \bv \bg^5 v, \quad    
A^{0} = \bp \bg^5 \bg^0 \p = \bv \bg^5 \bg^0 v, \nonumber \\
A^{3} &=& \bp \bg^5 \bg^3 \p = \bv \bg^5 \bg^3 v, \quad
V^{0} = \bp \bg^0 \p = \bv \bg^0 v, \quad
V^{3} = \bp \bg^3 \p = \bv \bg^3 v, \nonumber \\
Q^{30} &=& i \bp \bg^3 \bg^0 \p = i \bv \bg^3 \bg^0 v, \quad
Q^{21} = \bp  \bg^0 \bg^3 \bg^5 \p = 
i \bp  \bg^2 \bg^1 \p = i \bv  \bg^2 \bg^1 v, \nonumber 
\end{eqnarray}
obeying the following system of equations: 
\begin{mathletters}
\label{inv}
\begin{eqnarray}                                
\dot S_0 -2 \frac{k}{c} Q_{0}^{30} - 2 {\cG} A_{0}^{0} &=& 0, \\
\dot P_0 -2 \frac{k}{c} Q_{0}^{21} - 2 \Phi A_{0}^{0} &=& 0, \\                
\dot A_{0}^{0} -\frac{m-n}{c} A_{0}^{3} + 2 \Phi P_0 + 2 {\cG} S_0 &=& 0,\\
\dot A_{0}^{3} -\frac{m-n}{c} A_{0}^{0} &=& 0, \\                 
\dot V_{0}^{0} - \frac{m-n}{c} V_{0}^{3} &=& 0, \\                
\dot V_{0}^{3} - \frac{m-n}{c} V_{0}^{0} + 
2 \Phi Q_{0}^{30} - 2 {\cG} Q_{0}^{21} &=& 0,\\
\dot Q_{0}^{30} + 2\frac{k}{c}S_0 - 2 \Phi V_{0}^{3} &=& 0, \\                 
\dot Q_{0}^{21} +2 \frac{k}{c}P_0 + 2 {\cG} V_{0}^{3} &=& 0,                 
\end{eqnarray}
\end{mathletters}
where we denote $F_0 = \tau F$.
Combining these equations together and taking the first integral one gets
\begin{mathletters}
\label{fint}
\begin{eqnarray}                                
(S_{0})^{2} + (P_{0})^{2} + (A_{0}^{0})^{2} - (A_{0}^{3})^{2} +
(V_{0}^{0})^{2} + (V_{0}^{3})^{2} + 
(Q_{0}^{30})^{2} + (Q_{0}^{21})^{2} = C = {\rm Const} 
\label{I2}             
\end{eqnarray}
\end{mathletters}

Before dealing with the Einstein equations (\ref{ein}) let us go back to
(\ref{u}). From the first and the third equations of the system
(\ref{u}) one finds
\begin{equation}
\dot{u}_{13} = ({\cG} - Q) u_{13}^{2} - 2 i \Phi u_{13}
+ ({\cG} + Q),
\label{rik}
\end{equation}
where, we denote $u_{13} = u_1/u_3$ and $Q=[m - n - 2ik]/2c$.
The equation (\ref{rik}) is a Riccati one~\cite{Kamke} with variable
coefficients. Further, setting 
$u_{13} = v_{13} {\rm exp}[-2i \int \Phi(t) dt]$, from (\ref{rik})
one obtains
\begin{equation}
\dot{v}_{13} = ({\cG} - Q) v_{13}^{2} e^{-2i \int \Phi(t) dt}
+ ({\cG} + Q)e^{2i \int \Phi(t) dt}.
\label{v13}
\end{equation}        
The general solution of (\ref{v13}) can be written as
\begin{equation}
v_{13} = -\Biggl[\int ({\cG} - Q) e^{-2i \int \Phi(t) dt} dt + C(t)
\Biggr]^{-1},
%\label{}
\end{equation}     
with the integration constant be defined from
\begin{equation}
\int \Biggl[\int ({\cG} - Q) e^{-2i \int \Phi(t) dt} dt + C(t)
\Biggr]^{-2} dC = \int ({\cG} + Q) e^{2i \int \Phi(t) dt} dt.
%\label{}
\end{equation}
Thus given the nonlinear term in the Lagrangian and solution
of the Einstein equations, one finds the relation between 
$u_1$ and $u_3$ ($u_2$ and $u_4$ as well), hence the components
of the spinor field. 

Now we study the Einstein equation (\ref{ein}). In doing so,
we write the components of the energy-momentum tensor, which 
in our case read
\begin{mathletters}
\label{cemt}
\begin{eqnarray}
T_{0}^{0}&=& mS - F + \frac{k}{c} V^3, \\
T_{1}^{1} &=& T_{2}^{2} = {\cD} S + {\cG} P - F, \\  
T_{3}^{3}&=& {\cD} S + {\cG} P - F - \frac{k}{c} V^3, \\
T_{3}^{0} &=& - k V^0.  
\end{eqnarray}
\end{mathletters}
Let us demand the energy-momentum tensor to be conserved, i.e.,
\begin{equation}
T_{\nu;\mu}^{\mu} = T_{\nu,\mu}^{\mu} + \Gamma_{\beta \mu}^{\mu}\,
T_{\nu}^{\beta} - \Gamma_{\nu \mu}^{\beta}\,T_{\beta}^{\mu} = 0
\label{emc}
\end{equation}
Taking into account that $T_{\mu}^{\nu}$ is a function of $t$ only, from
(\ref{emc}) we find
\begin{mathletters}
\label{emcex}
\begin{eqnarray}
\Phi \dot S_{0} - {\cG} \dot P_{0} +\frac{k}{c} \dot V_{0}^{3}
- \frac{k}{c} \frac{m - n}{c} V_{0}^{0} &=& 0, \\
\dot V_{0}^{0} - \frac{m - n}{c} V_{0}^{3} &=& 0.
\end{eqnarray}
\end{mathletters}
As one can easily verify, the equations (\ref{emcex}) are 
consistent with those of (\ref{inv}). 

Let us now deal with the Einstein equations (\ref{ein}).
In view of (\ref{cemt}), from (\ref{03}) one obtains 
\begin{equation}
a^m b^{-n}/c^{m-n} = {\cal N} {\rm exp}[\kappa k \int V^0 dt], \quad
{\cal N} = {\rm const.},
\label{abcrel}
\end{equation}
whereas, subtraction of (\ref{22}) and (\ref{11}) leads to 
\begin{equation}
\frac{d}{d t} 
\Bigl[\tau \frac{d}{dt}\Bigl\{{\rm ln}\Bigl(\frac{a}{b}\Bigr)\Bigr\}\Bigr]
= \frac{m^2 - n^2}{c^2} \tau.
\label{ab1}
\end{equation}
Note that, the two other equations, obtained by subtracting 
(\ref{33}) from (\ref{11}) and (\ref{33}) from (\ref{22}), respectively,
are identical to (\ref{ab1}), that can be 
easily verified inserting (\ref{03}) and (\ref{inv}) into
the equations in question. Finally, summation of (\ref{11}), (\ref{22}),
(\ref{33}) and (\ref{00}), multiplied by 3, leads to the equation for
$\tau$, which in view of (\ref{cemt}) takes the form
\begin{equation}
\frac{\ddot \tau}{\tau} = 2 \frac{m^2 - mn - n^2}{c^2} - \frac{\kappa}{2}
\bigl[3(MS + {\cD} S + {\cG} P - 2 F) + 2 \frac{k}{c} V^3\bigr].
\label{detertau}
\end{equation}
For the right-hand side of the equation (\ref{detertau}) 
are some functions of $\tau$ only, the solution to this equation is 
well known~\cite{Kamke}. It can be shown that the quantities, related
to the spinor field are indeed the functions of $\tau$. Now, if we 
assume the metric function $c$ also to be a function of $\tau$,
both (\ref{ab1}) and (\ref{detertau}) can be solved explicitly.
Thus both the nonlinear spinor and the gravitational field equations
are solved in general. In what follows, we consider some special
cases and compare the solutions obtained with those for a BI
universe. In can be emphasized that the introduction of inhomogeneity
both in gravitational (through $m$ and $n$) and spinor (through $k$)
significantly complicates the whole picture.

To begin with we consider the linear case setting $F(I,J) = 0$.
As one sees, the structures of the spinor field equation (\ref{u}),
as well as the gravitational ones (\ref{ab1}) and (\ref{detertau})
in this case remain unaltered. Thus, the removal of the nonlinearity 
has little to offer in our cause. As it was mentioned earlier, the 
introduction of inhomogeneity is in the root of all these troubles. 
So, for some break-through we demand the spinor field completely
space-independent setting $k=0$. In this case for the components 
of the energy-momentum tensor immediately we find
\begin{mathletters}
\label{cemtk}
\begin{eqnarray}
T_{0}^{0}&=& mS - F, \\
T_{1}^{1} &=& T_{2}^{2} = T_{3}^{3} = {\cD} S + {\cG} P - F, \\  
T_{3}^{0} &=& 0.  
\end{eqnarray}
\end{mathletters}
In view of (\ref{cemtk}) from (\ref{abcrel}) we obtain
\begin{equation}
a^{m}b^{-n}/c^{m-n} = {\cal N}, 
\label{ac1}
\end{equation}
but the equations (\ref{u}), (\ref{ab1}) and (\ref{detertau}) are 
still unchanged. Thus we see that even the elimination of spinor
field inhomogeneity has little to offer. 

Finally, we consider the case, when the spinor field is independent
of space i.e., $k=0$ and the gravitational field is given by a 
BV space-time with $m = n$ in (\ref{met}). In this case for the spinor
field we find
\begin{mathletters}
\label{u5}
\begin{eqnarray} 
\dot{u}_{1} + i \Phi u_{1} - {\cG} u_{3} &=& 0, \\
\dot{u}_{2} + i \Phi u_{2} - {\cG} u_{4} &=& 0, \\
\dot{u}_{3} - i \Phi u_{3} + {\cG} u_{1} &=& 0, \\
\dot{u}_{4} - i \Phi u_{4} + {\cG} u_{2} &=& 0. 
\end{eqnarray} 
\end{mathletters}
The bilinear spinor forms in this case satisfy  
\begin{mathletters}
\label{inv5}
\begin{eqnarray}                                
\dot S_0 - 2 {\cG} A_{0}^{0} &=& 0, \\
\dot P_0 - 2 \Phi A_{0}^{0} &=& 0, \\                
\dot A_{0}^{0} + 2 \Phi P_0 + 2 {\cG} S_0 &=& 0,\\
\dot A_{0}^{3}  &=& 0, \\                 
\dot V_{0}^{0}  &=& 0, \\                
\dot V_{0}^{3} + 2 \Phi Q_{0}^{30} - 2 {\cG} Q_{0}^{21} &=& 0,\\
\dot Q_{0}^{30} - 2 \Phi V_{0}^{3} &=& 0, \\                 
\dot Q_{0}^{21} + 2 {\cG} V_{0}^{3} &=& 0,                 
\end{eqnarray}
\end{mathletters}
with the relations
\begin{mathletters}
\label{rel5}
\begin{eqnarray}
(S_{0})^{2} + (P_{0})^{2} + (A_{0}^{0})^2 &=& B_1,\\
A_{0}^{3} &=& B_2,\\
V_{0}^{0} &=& B_3,\\
(V_{0}^{3})^2 + (Q_{0}^{30})^2 + (Q_{0}^{21})^2 &=& B_4,
\end{eqnarray}
\end{mathletters}
with $B_i$ being the constant of integration.

For the gravitational field we find,
\begin{equation}
a = ({\cal N})^{(1/m)} b, 
\label{ab2}
\end{equation}
with the equations
\begin{equation}
\frac{d}{d t} 
\Bigl[\tau \frac{d}{dt}\Bigl\{{\rm ln}\Bigl(\frac{b}{c}\Bigr)\Bigr\}\Bigr]
= -\frac{2m^2}{c^2} \tau,
\label{ab3}
\end{equation}
and
\begin{equation}
\frac{\ddot \tau}{\tau} = -2 \frac{m^2}{c^2} - \frac{\kappa}{2}
\bigl[3(MS + {\cD} S + {\cG} P - 2 F)\bigr].
\label{detertau1}
\end{equation}
Note that the spinor field equation (\ref{u5}) completely coincides 
with those for a BI metric. 
If the nonlinear term in the Lagrangian is given as $F = F(I)$, then
the components of the spinor field can be given as
~\cite{sahaprd}
\begin{mathletters}
\label{psinl}
\begin{eqnarray} 
\psi_1(t) &=& (C_1/\sqrt{\tau}) {\rm exp}\,[-i\int(M - {\cD}) dt],\\
\psi_2(t) &=& (C_2/\sqrt{\tau}) {\rm exp}\,[-i\int(M - {\cD}) dt],\\
\psi_3(t) &=& (C_3/\sqrt{\tau}) {\rm exp}\,[i\int(M - {\cD}) dt],\\
\psi_4(t) &=& (C_4/\sqrt{\tau}) {\rm exp}\,[i\int(M - {\cD}) dt],
\end{eqnarray} 
\end{mathletters}
with $C_1,\,C_2,\,C_3,\,C_4$ being the integration constants, such that 
 $$C_{1}^{2} + C_{2}^{2} - C_{3}^{2} - C_{4}^{2} = C_0,$$
with $$ S = C_0/\tau.$$
In case, the nonlinear term is given by $F = F(J)$, the components
of the spinor field have the form
\begin{mathletters}
\label{psij}
\begin{eqnarray}
\psi_1 &=&\frac{1}{\sqrt{\tau}} \bigl(D_1 e^{i \sigma} + 
iD_3 e^{-i\sigma}\bigr), \\
\psi_2 &=&\frac{1}{\sqrt{\tau}} \bigl(D_2 e^{i \sigma} + 
iD_4 e^{-i\sigma}\bigr),  \\
\psi_3 &=&\frac{1}{\sqrt{\tau}} \bigl(iD_1 e^{i \sigma} + 
D_3 e^{-i \sigma}\bigr), \\
\psi_4 &=&\frac{1}{\sqrt{\tau}} \bigl(iD_2 e^{i \sigma} + 
D_4 e^{-i\sigma}\bigr).
\end{eqnarray} 
\end{mathletters}
Here $\sigma = int {\cG} dt$, and the integration constants
$D_i$ obey 
$$2\,(D_{1}^{2} + D_{2}^{2} - D_{3}^{2} -D_{4}^{2}) = D_0,$$
with $D_0$ to be determined from $$ P = D_0/\tau.$$ 

Contrary to the BI metric, where the metric functions $a,\,b,\,c$ and
$\tau$ are easily determined given the concrete form of nonlinearity,
in the case considered, the process is much complicated with 
the space inhomogeneity being an active player. Thus we see that
the introduction of inhomogeneity in the space-time has a far reaching
effect in the evolution of both spinor and gravitational field.


\noindent
\begin{thebibliography}{99}
\bibitem{ivanenko1} D. Ivanenko, Phys. Zs. Sowjetunion {\bf 13}, 141 (1938).
\bibitem{ivanenko2} D. Ivanenko, Soviet Physics Uspekhi {\bf 32}, 149 (1947).
\bibitem{rodichev} V. Rodichev, Soviet Physics JETP {\bf 13}, 1029 (1961).
\bibitem{sahapfu1} Yu.P. Rybakov, B. Saha and G.N. Shikin, 
PFU Reports: Physics {\bf 2}, (2), 61 (1994).
\bibitem{sahactp1} Yu.P. Rybakov, B. Saha and G.N. Shikin,
Communications in Theoretical Physics {\bf 3}, 199 (1994).
\bibitem{sahajmp} B. Saha and G.N. Shikin,
Journal Mathematical Physics {\bf 38}, 5305 (1997).
\bibitem{sahactp2} R. Alvarado, Yu.P. Rybakov, B. Saha and G.N. 
Shikin, JINR Preprint {\bf E2-95-16}, 11 p. (1995), 
Communications in Theoretical Physics {\bf 4}, (2), 247 (1995), 
gr-qc/9603035.
\bibitem{sahaizv} R. Alvarado, Yu.P. Rybakov, B. Saha and 
G.N. Shikin, Izvestia Vysshikh Uchebnikh Zavedenii: Fizika {\bf 38},
(7) 53 (1995) [Russ. Phys. J. {\bf 38}, 700 (1995)].
\bibitem{sahagrg} B. Saha, and G.N. Shikin, 
General Relativity and Gravitation {\bf 29}, 1099 (1997).
\bibitem{pfu-l} Bijan Saha, and G.N. Shikin, 
PFU Reports: Physics {\bf 8}, (1), 17 (2000); gr-qc/0102059.
\bibitem{sahal} Bijan Saha, Modern Physics Letters A {\bf 16}, 
1287 (2001); gr-qc/0009002.
\bibitem{sahaprd} Bijan Saha, Physical Review D {\bf 64}, 123501 (2001); 
gr-qc/0107013.
\bibitem{Ber}V.B.~Berestetski, E.M.~Lifshitz and L.P.~Pitaevski,
{\it Quantum Electrodynamics} (Nauka, Moscow, 1989).
\bibitem{Zelnor} V.A. Zhelnorovich, {\it Spinor theory and its 
application in physics and mechanics} (Nauka, Moscow, 1982).
\bibitem{Kamke} E. Kamke, {\it Differentialgleichungen losungsmethoden
und losungen} (Akademische Verlagsgesellschaft, Leipzig, 1957). 
\end{thebibliography}
\end{document}